\documentclass[12pt]{article}
\usepackage{latexsym}

\def\bs{\begin{subequations}}
\def\es{\end{subequations}}

\catcode`\@=11

\newtoks\@stequation
\def\subequations{\refstepcounter{equation}
  \edef\@savedequation{\the\c@equation}%
  \@stequation=\expandafter{\theequation}
  \edef\@savedtheequation{\the\@stequation}
  \edef\oldtheequation{\theequation}%
  \setcounter{equation}{0}%
  \def\theequation{\oldtheequation\alph{equation}}}

\def\endsubequations{\setcounter{equation}{\@savedequation}%
  \@stequation=\expandafter{\@savedtheequation}%
  \edef\theequation{\the\@stequation}\global\@ignoretrue}

\makeatletter
        \renewcommand{\theequation}{\thesection.\arabic{equation}}%
        \@addtoreset{equation}{section}%
\makeatother

\renewcommand{\thefootnote}{\fnsymbol{footnote}}

\begin{document}

\begin{titlepage}

 revised March 29, 2011

\begin{center}        \hfill   \\
            \hfill     \\
                                \hfill   \\

\vskip .25in

{\large \bf Tachyons in General Relativity \\}

\vskip 0.3in

Charles Schwartz\footnote{E-mail: schwartz@physics.berkeley.edu}

\vskip 0.15in

{\em Department of Physics,
     University of California\\
     Berkeley, California 94720}
        
\end{center}

\vskip .3in

\vfill

\begin{abstract}

We consider the motion of tachyons (faster-than-light particles) in the 
framework of General Relativity. An important feature is the large 
contribution of low energy tachyons to the energy-momentum tensor. 
We also calculate the gravitational field 
produced by tachyons in particular geometric arrangements; and it appears 
that there could be self-cohering bundles of such matter. This leads 
us to suggest that such theoretical ideas might be
relevant to major problems (dark matter and dark energy) in  
current cosmological models.

\end{abstract}

\vfill

\end{titlepage}

\renewcommand{\thefootnote}{\arabic{footnote}}
\setcounter{footnote}{0}
\renewcommand{\thepage}{\arabic{page}}
\setcounter{page}{1}

\section{Introduction}

While there is no reliable evidence for the existence of 
faster-than-light particles (tachyons) some physicists have  
considered them a theoretical possibility. \cite{ER1,ER2} Yet it appears that many 
carry a prejudice against the legitimacy of tachyonic 
theories because 
of concerns about causality.  This comes from  a familiar 
  ``paradox''  that posits communication via 
tachyons that appear to travel  
 backwards in time.  In Appendix A I have tried to dampen that 
prejudice by looking carefully at the assumptions behind the   
paradox.  If tachyons interact only weakly with ordinary matter, 
yet are profusely present throughout the universe, then it may be 
that those  objections become nugatory.

The present paper undertakes an investigation of how faster-than-light
particles might be 
considered within the General Theory of Relativity. There has been 
some earlier study of this \cite{ER3} but the present work covers 
considerable new ground and perhaps may make   
some interesting contributions to current models of cosmology. 

We start, in Section 2, with how they would be described in Special Relativity, 
noting the particular role of the energy-momentum tensor $T^{\mu 
\nu}$. Then we turn to General Relativity. In Sections 3 and 4 we study the 
motion of a tachyon in two familiar gravitational models. Then in 
Sections 5 and 6 we derive the gravitational field produced by a 
uniform flow of tachyons; and find that there is an attractive force 
on other nearby tachyons which increases in strength as their energy 
decreases. In Sections 7, 8, 9 and 10 we explore various models for 
 gravitational fields produced by static flows of 
tachyons, using the linear approximation to Einstein's equation; and 
we find that self-cohering bundles of tachyonic matter are feasible.
Section 11 has a note about how tachyons might contribute to 
gravitational lensing. These results, summarized in 
Section 12,  suggest that tachyons should be given closer attention in the  
ongoing study of cosmology.

\section{Special Relativity}

In Special Relativity, we have the motion of a particle described by 
the four-vector coordinate $\xi^{\mu}(\tau),\; \mu = 0,1,2,3$ and 
$\tau$ is a parameter called the proper time. For a free particle 
moving in a straight line with velocity $\textbf{v}$ we have,
\begin{equation}
\xi^{\mu}(\tau) = ( \gamma \tau, \gamma \textbf{v}\tau).\label{a1}
\end{equation}
Then, writing $\dot{\xi}^{\mu} = d\xi^{\mu}/d\tau$, we calculate
\begin{equation}
\eta_{\mu \nu}\;\dot{\xi}^{\mu}\;\dot{\xi}^{\nu} = \dot{\xi}^{\mu}\;\dot{\xi}_{\mu}
= \gamma^{2}\;(1-v^{2}), \label{a2}
\end{equation}
where we use the metric of Special Relativity, $\eta_{\mu \nu}$ (with 
$\eta_{00} = +1$ and the other diagonal components equal to -1) to 
raise and lower indices.
We define $\gamma = 1/\sqrt{|1-v^{2}|}$ so that this is always a real 
number. Then we see that for normal (slower-than-light) particles the 
invariant in 
Eq. (\ref{a2}) has the value +1 and for 
tachyons it is -1. (I am using units in which the velocity of light 
is $c=1.$)

Trying to follow conventional habits we can define the "energy'' and 
``momentum'' of a free tachyon as follows:
\begin{equation}
E = m\;\gamma, \;\;\;\;\;\; \textbf{p} = m\;\gamma\; \textbf{v}, 
\;\;\;\;\; E^{2} = p^{2} - m^{2}\label{a2a}
\end{equation}
where $m$ is a (real, positive) mass parameter.  This allows us to 
characterize the full set of free-particle tachyon states as,
\begin{equation}
E \ge 0, \;\;\;\;\; p^{2} \ge m^{2};\label{a2b}
\end{equation}
and the direction of the vector $\textbf{p}$ covers the entire sphere.

Troubles occur when one regards this $E$ and $\textbf{p}$ as 
a 4-vector. For tachyons this is a space-like 
vector and so the sign of the first component can change under a 
proper Lorentz transformation. This can lead to unnecessary confusion about 
negative energy states. This confusion was straightened out 
 in an earlier study \cite{CS}, where it was shown how to properly 
 use the labels for ``in'' and ``out'' states in any interaction. It 
 was also noted there that the 
best way to talk about energy and momentum is by starting with the 
familiar second rank tensor.

The energy-momentum tensor for a classical particle can be written as 
\begin{equation}
T^{\mu \nu}(x) = m \int\;d\tau\; 
\dot{\xi}^{\mu}\dot{\xi}^{\nu}\;\delta^{4}(x-\xi(\tau))\label{a3}
\end{equation}
and we note that the energy density, $T^{00}$, is always positive.
  For the 
free-particle motion noted above, we calculate that this tensor 
density has the diagonal elements $\rho, P_{11}, P_{22}, P_{33}$, 
where
\begin{equation}
\rho = m\gamma, \;\;\;\;\; P_{ii} = m\gamma v_{i}^{2}.\label{a4}
\end{equation}
Note that for tachyons the pressure terms ($\sum_{i}\;P_{ii}$) come 
out larger than 
the energy term ($\rho$); this is exactly the opposite of the 
situation for normal particles.\footnote{It is usually so that a repeated 
index implies summation; but there are exceptions, as seen here.}

To illuminate this unfamiliar situation, imagine the neutrino as  
being a tachyon. Then its contribution to the energy-momentum 
tensor, in the pressure term, would be larger than what one would 
ascribe to a massless particle by a factor of $v^{2} = (p/E)^{2}= (E^{2}+ 
m^{2})/E^{2}$. If one were to assume a mass of around 0.1 electron Volts and 
an average energy corresponding to the cosmic background temperature of around 
$3\;^{o}K$, then this would give a numerical factor of  $v^{2} 
\sim 10^{5}$. 
That would, presumably, have a significant impact on current 
calculations of cosmological models. \cite{SW}

For some further musings on kinematical aspects of tachyons, see 
Appendix B.

\section{General Relativity metric for uniform cosmology}

The Robertson-Walker metric, describing a homogeneous 
isotropic universe,  is 

\begin{equation}
\pm ds^{2} = dt^{2} -A(t)[B(r) dr^{2} + r^{2}d\theta^{2} + 
r^{2}sin^{2}\theta\;d\varphi^{2}].\label{b1}
\end{equation}
where we have identified the four (spherical polar) coordinates 
$t,r,\theta,\varphi$. The functions $A(t), B(r)$ are commonly written 
as:
\begin{equation}
A(t) = a(t)^{2}/c^{2}; \;\;\;\;\; B(r) = \frac{1}{1-\epsilon 
r^{2}/R^{2}}.\label{b2}
\end{equation}

With this metric we construct the following equations for the geodesic.
\begin{eqnarray}
\ddot{t}  + \frac{1}{2}A^{\prime}[\dot{r}^{2} B + 
\dot{\theta}^{2}r^{2} + \dot{\varphi}^{2}r^{2}sin^{2}\theta]=0,\label{b3}\\
-AB \ddot{r} + \dot{r}[-\dot{t}A^{\prime}B - \dot{r} 
AB^{\prime}] + \frac{1}{2}A[\dot{r}^{2}B^{\prime} 
+\dot{\theta}^{2}2r + \dot{\varphi}^{2}2r\;sin^{2}\theta] = 0,\label{b4} \\
-Ar^{2}\ddot{\theta} 
+\dot{\theta}[-\dot{t}A^{\prime}r^{2}-\dot{r}2Ar] + 
\frac{1}{2}A[\dot{\varphi}^{2}r^{2}2\;sin\theta\;cos\theta] = 0,\label{b5} \\
-Ar^{2}\;sin^{2}\theta\; \ddot{\varphi} + \dot{\varphi}[-\dot{t} 
A^{\prime}r^{2}\;sin^{2}\theta -\dot{r}A2r\;sin^{2}\theta 
-\dot{\theta}Ar^{2}2\;sin\theta\;cos\theta] = 0,\label{b6}
\end{eqnarray}
where the ``prime'' on $A$ or $B$ means derivative with respect 
to its argument.

We can integrate the last two equations, yielding 
\begin{eqnarray}
Ar^{2}\;sin^{2}\theta\;\dot{\varphi} = M, \label{b7}\\
A^{2}r^{4}\dot{\theta}^{2} + M^{2}/sin^{2}\theta  = L^{2},\label{b8}
\end{eqnarray}
where M and L are some constants.  Then, substituting these results 
back into the first two equations we also get the integrals:
\begin{eqnarray}
A^{2}B \dot{r}^{2} + L^{2}/r^{2} = 2{\cal{E}}, \label{b9}\\
\dot{t}^{2} -2{\cal{E}}/A = Q,\label{b10}
\end{eqnarray}
with two more constants, ${\cal{E}}$ and Q.

Using these results we now calculate the invariant 
\begin{equation}
g_{\mu \nu}\;\dot{x}^{\mu}\;\dot{x}^{\nu} = Q,\label{b11}
\end{equation}
 So, for normal (slower-than-light) 
particles we would choose $Q=+1$; for light we would choose $Q=0$; 
and for tachyons, we choose $Q=-1$. We shall see that in some 
instances the constant $2{\cal{E}}$ may be taken as $v^{2}$ for some initial 
velocity $v$, while L and M relate to the angular momentum for unit 
mass.

From Eq. (\ref{b9}) we see that motion of a tachyon will be 
contained in the case of a closed universe, ($\epsilon = +1$), even 
though its velocity may be arbitrarily large. This 
follows because the factor $B(r)$ becomes infinite at $r=R$, and thus 
the radial coordinate $r$ must stop its increase at that point.

Another simple solution of Eq. (\ref{b9}) is a circular orbit 
($\dot{r} = 0$) at a radius $r_{0} = L/v$. But such orbits are not 
stable: for $B>0$ they will spiral outward; and they will spiral 
inward for $B<0$, which may occur for $r_{0} > R$ in a closed 
universe.

\section{Tachyon in Schwarzschild metric}

Outside of a central source of gravitation, we have the Schwarzschild 
metric,
\begin{equation}
\pm ds^{2} = A(r)dt^{2} -[A(r)^{-1} dr^{2} + r^{2}d\theta^{2} + 
r^{2}sin^{2}\theta\;d\varphi^{2}].\label{c1}
\end{equation}
where $A(r) = (1-r_{s}/r)$ and $r_{s} = 2GM$.

The solution of the geodesic equations in this metric yields 
the following.
\begin{equation}
r^{2}\dot{\varphi} = L, \;\;\;\;\; A \dot{t} = \gamma, \;\;\;\;\; 
A^{-1}(\dot{r}^{2} -\gamma^{2})+ L^{2}/r^{2} = -Q \label{c2}
\end{equation}
where $\gamma$ and L are constants of integration and $Q$ is the same 
constant (with values +1, 0, -1) defined in Eq. (\ref{b11}).

For slower-than-light particles (Q=+1) this gives  the 
Kepler orbits for $r > r_{s}$, along with small relativistic 
corrections. For tachyons (Q=-1) we see that this
does not give localized orbits but only scattering states.

Introducing the variable, $u = 1/r(\varphi)$, and introducing the 
asymptotic velocity $v$, with $\gamma^{2} = Q + v^{2}\gamma^{2}$, $L 
= bv\gamma$, we get the 
equation for the orbital,
\begin{equation}
u^{\prime \;2} + u^{2}  - \frac{ 
Q r_{s} u +v^{2}\gamma^{2}}{L^{2}} =r_{s}u^{3}.\label{c3}
\end{equation}

From this equation, we can calculate the scattering angle to first order in the 
gravitational field strength and find the result
\begin{equation}
\Delta \phi =\frac{r_{s}}{b}\;\chi, \;\;\;\;\; \chi =[2 + 
\frac{Q}{v^{2}\gamma^{2}}].
\label{c4}
\end{equation}
For light (Q=0), we have $\chi = 2$, which is the well known result. For 
tachyons (Q=-1), $\chi$ has the value 2 at high energies and then 
decreases to 1 as the energy drops. 
 For ordinary particles (Q=+1)  at very high energies (v 
close to 1) $\chi$ is again 2; 
and at much lower velocities we should not accept the first order 
approximation but solve the exact Kepler problem.

We can also easily look at a ``head-on'' collision by setting $L=0$ 
in Eq. (\ref{c2}). Then we 
see how a tachyon will behave approaching the edge of a black hole, 
$r \rightarrow r_{s}$.

\section{Metric in cylindrical symmetry}

For  tachyons as a gravitational source, the simplest model is 
 a straight line of particles, flowing constantly along the z-axis, 
with equal velocities in both directions.  
With cylindrical symmetry we use
the four coordinates $\xi^{\mu} = (t, r, \theta, z)$, 
where $r = \sqrt{x^{2}+y^{2}}$.  The source 
is then a diagonal energy-momentum tensor density, following Eq. 
(\ref{a4}),

\begin{equation}
T^{00} = \rho \gamma \delta(r)/(2\pi r), \;\;\;\;\;T^{11} = T^{22} = 
0, \;\;\;\;\;
T^{33} = \rho \gamma v^{2} \delta(r)/(2\pi r),\label{e0} 
\end{equation}
where $\rho$ is the mass per unit length.
We now make this ansatz for the metric:

\begin{equation}
\pm ds^{2} = A(r) dt^{2} -B(r) dr^{2} - r^{2}d\theta ^{2} - C(r) 
dz^{2}.\label{e1}
\end{equation}

Next, we calculate the elements of the Christoffel symbol.
\begin{eqnarray}
\Gamma^{0}_{01} = \Gamma^{0}_{10} = \frac{A'}{2A} \\
\Gamma^{1}_{00} = \frac{A'}{2B}, \;\;\;\;\; \Gamma^{1}_{11} = 
\frac{B'}{2B} \\
\Gamma^{1}_{22} = -r/B, \;\;\;\;\; \Gamma^{1}_{33}=- \frac{C'}{2B} \\
\Gamma^{2}_{12}=\Gamma^{2}_{21} = 1/r, \;\;\;\;\; 
\Gamma^{3}_{13} = \Gamma^{3}_{31} = \frac{C'}{2C}, \label{e2}
\end{eqnarray}
where the prime means derivative with respect to $r$.

From this, we calculate the components of the Einstein tensor 
$G_{\mu\nu} = R_{\mu \nu} - g_{\mu \nu} R/2$.
\begin{eqnarray}
G_{00} = \frac{A}{2B} \{\frac{C''}{C} -\frac{C'^{2}}{2C^{2}}-
\frac{B'C'}{2BC} + \frac{C'}{rC} -\frac{B'}{rB}\} \\
G_{11} = -\frac{C'}{2rC} - \frac{A'}{2rA}-\frac{A'C'}{4AC} \\
G_{22} =\frac{r^{2}}{2B}\{-\frac{A''}{A}+\frac{A'^{2}}{2A^{2}}+ 
\frac{A'B'}{2AB} - \frac{C''}{C}+\frac{C'^{2}}{2C^{2}} - 
\frac{A'C'}{2AC} +\frac{B'C'}{2BC}  \} \\
G_{33} = \frac{C}{2B}\{-\frac{A''}{A}+\frac{A'^{2}}{2A^{2}} 
+\frac{A'B'}{2AB} -\frac{A'}{rA}+\frac{B'}{rB}\}.\label{e3}
\end{eqnarray}

If we set all these components equal to zero, we find the solutions
\begin{eqnarray}
A = a \;r^{\alpha}, \;\;\;\;\; B = b\;r^{\beta} \;\;\;\;\; 
C = c\;r^{\gamma} \\ 
\beta = \frac{\alpha^{2}}{\alpha+2}, \;\;\;\gamma = 
\frac{-2\alpha}{\alpha+2}\label{e4}
\end{eqnarray}
where $\alpha$ is some constant ($\neq -2$).  

For small $\alpha$ we have the 
(``weak field'') solutions:
$A = 1+\alpha \;ln(r)$, $B = 1$, $C = 1-\alpha\; ln(r)$.

Now, using the Einstein field equation $G_{\mu \nu} = -8 \pi GT_{\mu 
\nu}$, with Eq. (\ref{e0}) we find 
 $\alpha/2 = 8\pi G \rho \gamma /(2\pi)$. This leads to the result
\begin{equation}
Force = -\Gamma^{1}_{00} = -4G\rho \gamma /r \label{e5}
\end{equation}
for the case of a slow particle being attracted by a line of 
mass-density $\rho$.  This is to be compared
with the result of Newtonian gravity; and there is a difference of a 
factor of 2. (This problem in General Relativity was studied long ago by 
Levi-Civita \cite{TLC} and he obtained 
the correct Newtonian force.) I have no explanation of this 
discrepancy. Maybe it 
has to do with the line of matter extending to infinity. 

Similarly, for a fast particle near a line of streaming tachyons  we 
find the attractive force
\begin{equation}
Force = -\Gamma^{1}_{33} \dot{\xi}^{3}\dot{\xi}^{3}= 
-4G\rho \gamma v^{2} (\dot{z}^{2})/r.\label{e6}
\end{equation}
What is novel here is that the strength of the gravitational 
attraction between tachyons increases as the velocity increases 
(without limit).

There is some question about the legitimacy of this result due to the 
non-linearity of the equations and the singularity at $r=0$. Please 
see Appendix C for resolution of this question.

\section{Circular model}

To make a more realistic model of tachyons as a gravitational 
source, we can imagine something like a circular flow, which is 
limited in its spatial extent. The full solution of Einstein's field 
equations then becomes more difficult; but we can start out by making 
the weak field approximation. The general approach (see, for example,  
\cite{SW1}) goes like this:
\begin{eqnarray}
(\frac{\partial}{\partial t}^{2} - \nabla ^{2})h_{\mu \nu} = 
-16 \pi G T_{\mu \nu}, \label{g1a}\\  
g_{\mu \nu} = \eta_{\mu \nu} +h_{\mu \nu}- \frac{1}{2} \eta_{\mu \nu} 
h, \;\;\;\;\; h = \eta^{\mu \nu}\;h_{\mu \nu},\label{g1b}
\end{eqnarray}
with the side condition (gauge choice)
\begin{equation}
\partial^{\mu}h_{\mu \nu} = 0.\label{g1c}
\end{equation}

This gives us Newtonian gravity in the familiar case of mass at rest 
($T_{00} = \rho (\textbf{x})$); and we use this now for a tachyon 
source, following (\ref{a3}), (\ref{a4}).

If this is a large uniform and static circulating flow of tachyons 
and we look close to the edge of it, 
then we can approximate it as an infinite cylinder, along the local 
z-axis, and immediately 
see the result. From $h_{33}$ we get an attractive force  on a unit mass at 
rest; and it is  without that previous extra factor of 2.
\begin{equation}
F = -2G P/r\label{g2}
\end{equation}
where r is the distance from the center line of this cylinder and $P$ 
is the ``pressure'' ($m\gamma v^{2}$) per unit length of the source.
For large values of $v$, which means low energy for the tachyons in 
the source, 
this force becomes large, proportional to $|v|$ (or $1/E$). This 
suggests remarkable new physics, if such things actually exist.

Let us see how this force works upon another tachyon travelling nearby.
The geodesic equation gives us
\begin{equation}
\frac{d^{2} r}{d \tau^{2}} = F, \;\;\;\;\; \frac{d^{2}r}{dt^{2}} = 
F/\gamma^{2}\label{g3}
\end{equation}
where this $\gamma$ describes the free-moving tachyon. So,  a low 
energy tachyon is accelerated toward the flow of other low energy 
tachyons. This effect increases with the velocity of each component 
and suggests how such a structure - a self sustaining and accreting 
mass flow of tachyons held together by gravity - might evolve.

For example, if we have a total mass M of tachyons moving within a 
tube of radius $r$ at large velocity V 
in a circle of radius R, then an additional tachyon moving 
with large velocity along the edge of this tube  
will have its trajectory bent to keep up with that circle provided 
that $GMV/r > \pi$.

In Section 10 we shall carry this study further to include tachyon 
orbits spiraling  around the center line of this source.

Any such structure would also have gravitational effects upon ordinary 
matter and upon light beams.  This is what the geometric theory of 
gravity requires and we now try to consider this in some generality.

\section{Static gravitational fields}

Now we want to follow the first-order approach of Eqs. (\ref{g1a} - 
\ref{g1c}) for a general 
distribution of ordinary matter and tachyons that is static, not 
changing in time. We would like to be able to specify a  
source $T_{\mu \nu}(\textbf{x})$, for example, one that has a 
simple structure like that of Eq. (\ref{a4}), where the energy 
density $\rho$ and the pressure terms
$P_{i i}$ are now limited by some spatial envelope we might choose.

This model has an obvious shortcoming in that $\partial^{\mu}T_{\mu 
\nu}$ is not zero. (It would be zero if we only had the energy density and 
ignored the pressure terms; but for tachyons the motion is the main 
thing.) At first, I thought to put that criticism aside 
with the following argument.
The particles that comprise this source experience some acceleration in 
order to remain in a confined region and we assume that the requisite 
forces are themselves gravitational. Thus the error introduced into 
the Einstein equation will be of second order in the constant $G$ 
and we shall ignore it. (This seems analogous to the 
distinction between the ordinary derivative of $T_{\mu \nu}$ and its 
covariant derivative.) 

This argument, however, is false on two fronts.  In terms of the mathematics: The Einstein 
equation $G_{\mu \nu} = -8\pi G T_{\mu \nu}$ has, by construction, a 
left hand side that automatically satisfies the conservation law. 
Thus, using a right hand side that fails this test may (will) lead to 
incorrect solutions.  In terms of the physics: For non-relativistic 
particles moving in  gravitational orbits we know that the average 
potential energy is just as big as the kinetic energy. Therefore, it 
would be wrong to ignore the former and only keep the latter in 
constructing the energy-momentum tensor. (Again, for non-relativistic 
particles these pressure terms are generally negligible, of order 
$v^{2}/c^{2}$ compared to their rest-mass; but for tachyons they dominate.)

First, we shall find a general statement about the gravitational 
fields produced (in first order) from a  static source that is 
localized in space. Then we shall proceed to some particular models.

We have both $T_{\mu \nu}$ and $h_{\mu \nu}$ 
independent of the time variable, with the former assumed confined in 
some 
region of space. We also take the time-space components of each of 
these tensors to be zero.  The basic equations are,
\begin{eqnarray}
- \nabla^{2} h_{\mu \nu}(\textbf{x}) = -16 \pi G T_{\mu \nu}(\textbf{x}) 
\label{h1}\\
h_{00}(\textbf{x}) = -4 G \int d^{3}x' \;T_{00}(\textbf{x}') 
/|\textbf{x} - \textbf{x}'| \label{h2}\\
h_{ij}(\textbf{x}) = -4 G \int d^{3}x' \;T_{ij}(\textbf{x}') 
/|\textbf{x} - \textbf{x}'| \label{h3}\\
\partial_{i}h_{ij}(\textbf{x}) = \partial_{i}T_{ij}(\textbf{x}) = 
0,\label{h4}
\end{eqnarray}
with $i,j = 1,2,3$.
The asymptotic form of Eq. (\ref{h3}) is
\begin{equation}
h_{ij} \sim -\frac{4G}{r} \int d^{3}x' T_{ij}(\textbf{x}'),\label{h5}
\end{equation}
but from Eq. (\ref{h4}) we calculate,
\begin{equation}
0 = \int d^{3}x\;x_{k}\;\partial_{i}T_{ij}(\textbf{x}) = - \int 
d^{3}x\;T_{kj}(\textbf{x}).\label{h6}
\end{equation}

This means that we get no long range force from the spatial parts of 
the source, as suggested by Eq. (\ref{h5}); that comes only from the $T_{00}$
component, Eq. (\ref{h2}).  Is this 
well-known?  For ordinary (slow) matter this is no big deal (maybe it 
gets involved when one asks about $v^{2}/c^{2}$ corrections to 
Newtonian gravity). But for tachyons it would be the major feature of 
their gravitational role. If there are such static, localized 
configurations of tachyons as we have been speculating here, then 
they may make major contributions to the local gravitational field; 
but they produce little or no long range gravitational attraction. 
What a lovely way to explain both dark matter and dark energy!

Incidentally, Eq. (\ref{h6}) may be recognized as a generalization 
of the Virial Theorem, which is well known in non-relativistic 
mechanics, either the classical or the quantum variety.

What can one say about the next term, after $\sim 1/r$ in the 
asymptotic expansion of Eq.(\ref{h3})? It would appear to be $\sim 
1/r^{2}$ with coefficients $\int d^{3}x\;x_{k}\;T_{ij}(\textbf{x})$. 
These can be shown to vanish, however, by considering the identities 
$\int d^{3}x\;x_{k}\;x_{l}\;\partial_{i}T_{ij}=0$. The leading term 
will then be of order $1/r^{3}$, which is what one expects of a 
quadrupole source.

In what follows we shall try to construct models of the 
energy-momentum tensor that obey the conservation law, are static in 
time, and localized in space. That means they need to involve two parts which 
we may call 
the ``kinetic energy'' part and the ``potential energy'' part.

\section{Circular flows} 

We shall be interested in problems with 3-dimensional and also 
2-dimensional symmetry. So let us do this analysis first for 
n-dimensional Euclidean tensors $T_{i j}(\textbf{x})$, with $i,j = 1, 
\ldots n$. We have the radial coordinate $r = \sqrt{\sum_{i}x_{i}x_{i}}$.

For circular motion  we can see that the kinetic part 
of the nxn matrix $T_{ij}$ should look like this
\begin{equation}
T^{(kin)}_{ij} = (\delta_{ij}-x_{i}x_{j}/r^{2})b(r)  \sim 
(v_{i}\;v_{j})\label{i1}
\end{equation}
where any multiplying envelope $b(r)$ should be non-negative.

The easiest way to verify that this does represent circular flow is 
to calculate $x_{i}T^{(kin)}_{ij}$ and see that it is zero, since
there should be no velocity in the radial direction.

It is then convenient to introduce the nxn coordinate matrices $I$ and 
$D$ as follows .
\begin{eqnarray}
I_{ij} = \delta_{ij}, \;\;\;\;\; D_{ij} = \delta_{ij} - n 
x_{i}x_{j}/r^{2},\;\;\;\;\;\;\;\;\;\;\label{i2} \\
Trace(I) = n, \;\;\;\;\; Trace(D) = 0, \;\;\;\;\; D^{2} = (n-1)I - 
(n-2)D.\label{i3}
\end{eqnarray}

Then, we construct the full tensor as
\begin{eqnarray}
T_{i j} = \delta_{i j}a(r) + D_{i j} b(r) =\label{i4} \\
\delta_{i j}[a(r)-(n-1)b(r)] + nT^{(kin)}_{ij}.\label{i5}
\end{eqnarray}
and thus we identify $[a-(n-1)b]$ as the ``potential 
energy'' portion of $T$.  From the conservation requirement 
$\partial_{i}T_{ij} = 0$, we get the constraint 
\begin{equation}
r^{n}a'(r) = (n-1)(r^{n}b(r))', \;\;\;\;\; \frac{d}{dr} [a-(n-1)b] = 
n(n-1)b(r)/r \ge 0.\label{i6}
\end{equation}
Thus the part we identify as potential energy in the tensor is 
diagonal and increases with r.  Assuming that both the functions $a(r)$ 
and $b(r)$ go to zero at large $r$, this means that the potential 
energy is negative - an ``attractive well.''
From this we also learn that $\int_{0}^{\infty} r^{n-1} dr \;a(r) = 0$ .

One more general result is the following, regarding the 
n-dimensional Lapacian operator.
\begin{equation}
\nabla^{2}\;D\;g(r) = D [ g'' + (n-1)g'/r - 2n g/r^{2}].\label{i7}
\end{equation} 
	
Then, if we write
\begin{eqnarray}
h_{ij} = \delta_{ij}f(r) + D_{ij}g(r) \\
r^{n}f'(r) = (n-1)(r^{n}g(r))',
\end{eqnarray}
where we have required $\partial_{i}h_{ij}=0$, then we get the 
solution
\begin{eqnarray}
g(r) = -\frac{16\pi G}{n+2} \;[r^{-n}\int_{0}^{r}s^{n+1}ds\;b(s) + 
r^{2}\int_{r}^{\infty}s^{-1} ds\;b(s)]\label{i8} \\ 
f'(r) = -16\pi G(n-1)\;r \int_{r}^{\infty} s^{-1} ds\; b(s).\label{i9}
\end{eqnarray}
There are alternative solutions 
in terms of the source function 
$a(r)$; but they are determined by the relations stated above between 
$f$ and $g$, on the one hand, and $a$ and $b$ on the other.

From the fact that we have $b(r) \ge 0$, we can describe the two 
functions qualitatively as follows: $g(r)$ is everywhere negative and 
goes to zero at both $r \rightarrow 0$ and $r\rightarrow \infty$; 
$f(r)$ has everywhere a negative 
derivative and falls off to zero as r increases, thus it must be 
positive at all finite r.

The asymptotic solution at large $r$ is, as expected,  
\begin{equation}
g(r) \sim -\frac{16 \pi 
G}{(n+2)}\;r^{-n}\int_{0}^{\infty}s^{n+1}ds\;b(s),\label{i10}
\end{equation}
and the function $f(r)$ falls off even faster.

From the equations above one can derive  a number of identities, for 
example,
\begin{equation}
f(0) = -n(n-1) \int_{0}^{\infty} dr\; r^{-1}\;g(r) = 16\pi G 
\frac{n-1}{2}\int_{0}^{\infty}dr\;r\;b(r).\label{i11}
\end{equation}

Finally, that portion of the metric which depends on the function 
$g(r)$ can be simplified, as follows.
\begin{eqnarray}
\sum_{i, j = 1}^{n}\;dx_{i}\;dx_{j} [\delta_{ij} - n 
x_{i}x_{j}/r^{2}]\;g(r) = \label{i12}\\ 
\sum_{i}(dx_{i})^{2}\; g(r) - n \left( 
\sum_{i}x_{i}\;dx_{i}\right)^{2}\;g(r)/r^{2} 
=\label{i13} \\
\left(\sum_{i}(dx_{i})^{2}  - n (dr)^{2}\right)\;g(r). \label{i14}
\end{eqnarray}

\section{Spherical source}

We start by applying the machinery of the previous Section to the case of a 
spherically symmetric source, $n=3$. We have the following solution of 
the field equations, with $r = \sqrt{x^{2}+y^{2}+z^{2}}$,
\begin{equation}
g(r) = 
 -\frac{16}{5}\pi G [r^{-3}\int_{0}^{r}s^{4}ds\;b(s) + 
r^{2}\int_{r}^{\infty}s^{-1} ds\;b(s)]
 \sim -\frac{4G}{5r^{3}}\int d^{3}x\;r^{2}b(r),\label{j1}
\end{equation}
and the function $f(r)$ falls off even faster at large r.

We still have 
\begin{equation}
h_{00} = -4G\int d^{3}x' \frac{T_{00}(\textbf{x}')}{|\textbf{x} - 
\textbf{x}'|} \equiv 4V_{0}(r) \sim - r^{-1}.\label{j2}
\end{equation}

Next, we re-write the metric, changing from Cartesian to  spherical 
polar coordinates.
\begin{eqnarray}
ds^{2} = [1+A]dt^{2} + [-1+B]dr^{2} + [-1+C]\;r^{2}[d\theta^{2} + 
sin^{2}\theta\; d\varphi^{2}]\label{j3} \\ 
A(r) =2V_{0}+\frac{3}{2}f, \;\;\; B(r)=2V_{0}-\frac{1}{2}f - 2g, 
\;\;\; C(r) = 2V_{0}-\frac{1}{2}f + g.\label{j4}
\end{eqnarray}

Then, we  calculate the Christoffel symbols and put them into the 
geodesic equations. We immediately integrate two of them,
\begin{equation}
\dot{t} = \gamma e^{-A}, \;\;\;\;\; r^{2}\dot{\varphi} = L 
e^{C},\label{j5}
\end{equation} 
where $\gamma$ and $L$ are constants and, as usual, we choose $\theta = \pi/2$. 
Now the radial equation, 
in strictly first order form, leads to
\begin{equation}
(\dot{r})^{2} + \gamma^{2}A(r) + L^{2}/r^{2} = 2{\cal{E}} = 
\gamma^{2} -Q.\label{j6}
\end{equation}

The critical question we want to answer is whether tachyons may find 
bound orbits in this configuration of circulating tachyons (plus 
stationary matter). The potential $V_{0}$ is negative, falling off to 
zero at large r; and the additional component of $A(r)$, which is 
$f(r)$, merely adds to the familiar $L^{2}/r^{2}$ term. Thus, 
with $Q = -1$ for tachyons, we see that ${\cal{E}} >0$ and so 
conclude that this equation does \emph{not} 
allow bound states.

Since we are seeking a self-supporting arrangement of tachyons, held 
together by their own gravitational forces,  we must look for 
some other geometry.  That is where we turn now.

\section{Cylindrical source}

We now apply the above machinery  to the case of a 
cylindrically symmetric source, $n=2$. Here we have $T_{3i} = h_{3i} 
= 0$; and we have the following solution of 
the field equations, with $r = \sqrt{x^{2}+y^{2}}$:
\begin{equation}
g(r) =  -\frac{16}{4}\pi G [r^{-2}\int_{0}^{r}s^{3}ds\;b(s) + 
r^{2}\int_{r}^{\infty}s^{-1} ds\;b(s)]
 \sim -\frac{2G}{r^{2}}\int d^{2}x\;r^{2}b(r),\label{k1}
\end{equation}
and the function $f(r)$ falls off even faster at large r.

We still have 
\begin{eqnarray}
h_{00} = -4G\int d^{3}x' \frac{T_{00}(\textbf{x}')}{|\textbf{x} - 
\textbf{x}'|} \equiv 4V_{0}(r) \sim ln(r/R)  ,\label{k2}\\
h_{33} = -4G\int d^{3}x' \frac{T_{33}(\textbf{x}')}{|\textbf{x} - 
\textbf{x}'|}\equiv 4V_{3}(r) \sim ln(r/R)  ,\label{k3}
\end{eqnarray}
where $R$ is some measure of the linear extent of this source; in 
other words, these potentials are negative with a positive slope.

Next, we want to calculate the Christoffel symbols and put them into the 
geodesic equation. Changing from Cartesian coordinates to cylindrical 
coordinates $(t,r,\theta,z)$, we find,
\begin{eqnarray}
ds^{2} = (1+A)dt^{2} + (-1+B)dr^{2}+ (-1+C)r^{2}d \theta^{2} + 
(-1+D)dz^{2}\label{k4} \\ 
A(r) = [2V_{0}+2V_{3}+f], \;\;\;\;\; B(r)= [2V_{0}-2V_{3}-g],\label{k5} \\ 
C(r) = [2V_{0}-2V_{3}+g], \;\;\;\;\; D(r)=[2V_{0}+2V_{3}-f]. \label{k6}
\end{eqnarray}

The non-zero Christoffel symbols, to first order, are then
\begin {eqnarray}
\Gamma^{t}_{t r} = \Gamma^{r}_{t t} = \frac{1}{2}A' ,\label{k7}\\ 
\Gamma^{r}_{r r} =  -\frac{1}{2}B', \label{k8}\\ 
\Gamma^{\theta}_{ \theta r} = \frac{1}{r} -\frac{C'}{2}, \label{k9} \\
\Gamma^{r}_{\theta \theta} =  -r(1+B-C) + \frac{r^{2}C'}{2} ,\label{k10} \\ 
\Gamma^{z}_{z r} = -\Gamma^{r}_{z z} = - \frac{1}{2}D'.\label{k11}
\end{eqnarray}

Three of the four geodesic equations are immediately integrated, as 
follows,
\begin{equation}
\dot{t} = \gamma e^{-A}, \;\;\; \dot{z} = \kappa e^{D}, \;\;\; 
r^{2}\dot\theta = Le^{C},\label{k12}
\end{equation}
where $\gamma, \kappa, L$ are constants, which might be named energy, 
linear momentum and angular momentum, respectively.

The fourth equation, for the radial coordinate, is more complicated. 
In a strictly first-order approximation it can be integrated to
\begin{equation}
(\dot{r})^{2} + L^{2}/r^{2}+ \gamma^{2}A + \kappa^{2}D = 2 {\cal{E}} 
= \gamma^{2}-\kappa^{2}-Q,\label{k13}
\end{equation}
where, as always, $Q=-1$ for tachyons.

The question we address now is whether this equation allows bound 
orbits, which would imply particles moving in spirals around the 
z-axis. The effective potential in this radial equation is,
\begin{equation}
\frac{L^{2}}{2r^{2}} + U(r), \;\;\;\;\; U(r) = 
(\gamma^{2}+\kappa^{2})(V_{0}(r) + V_{3}(r)) 
+(\gamma^{2}-\kappa^{2})f(r)/2.\label{k14}
\end{equation}
For bound states we would require, at least, that $U(r)$ be a 
function that increases with r at large r. Both potentials $V_{0}$ 
and $V_{3}$ are of this character; and the function $f(r)$ falls off 
faster at large r.  So this requirement is satisfied. (This is no 
trivial result. It depends on how various plus-and-minus signs have 
worked out in this calculation.)  The next requirement for bound 
states is that the integration constant ${\cal{E}}$ have a value 
equal to or greater than the minimum point of the total effective 
potential.

To get an idea about this, we can say that for a free tachyon, or one 
in a scattering state,  we would
have,
\begin{equation}
\gamma^{2} = 1/(v^{2}-1), \;\;\; \kappa^{2} = 
\gamma^{2}\;v_{z}^{2},\label{k15}
\end{equation}
which leads to $2{\cal{E}} = \gamma^{2}(v^{2}-v_{z}^{2}) > 0$.

However, the integration parameters $\gamma$ and $\kappa$ are free 
from such constraints, and we see that  ${\cal{E}}$ certainly can be 
negative, thus allowing for bound state solutions. Note the contrast 
between this result and that of the previous Section, which looked at 
spherical symmetry.

This result expands on what we had seen in the earlier work of 
Sections 5 and 6, allowing for the circulatory motion in the source, 
along with the translational motion, relative to the z-axis.  We 
conclude that streams of tachyons could form into coherent localized 
bundles held together by their own gravitational forces.

We are now in the position to invite speculation about detailed 
clustering of large numbers of tachyons (assuming that such things exist) in 
something akin to rope-like structures throughout the universe. How 
big or small
might such ropes be - as tight bundles or wide tubes; how might they  
connect to themselves - in simple circles or in 
knots, in doughnuts or in halos;  and how might they  attach themselves, 
gravitationally, to 
large masses of ordinary matter? Model-making and detailed 
calculations are needed to work out the answers to these questions; 
and further expert analysis is needed to see how such models may 
comport with known observational data. 

\section{Gravitational lensing} 

The deflection of light beams by the gravitational field of dense 
matter is an important tool in the observation of galactic structures; 
and it has been used to impute the distribution of dark matter. If 
tachyonic matter actually exists, and is collected in static bundles, 
as has been suggested in the analysis of this paper, then one would 
want to ask how that might contribute to gravitational lensing. 

In the standard analysis, one works from the assumption that the source 
of gravitational fields is slow moving massive matter; and so one 
starts with the Newtonian potential, which we have called 
$V_{0}(\textbf{x})$ and  carries out a sort of ray-tracing 
calculation to get the pattern of light beam deflection.

If tachyonic matter is a significant presence, then this calculation 
needs to be augmented. It turns out that one should use, 
\begin{equation}
V_{0}(\textbf{x}) + V_{3}(\textbf{x}) = -G \int d^{3}x' \frac{1}
{|\textbf{x}-\textbf{x}'|}
[T_{00}(\textbf{x}') + T_{33}(\textbf{x}')],\label{l1}
\end{equation}
where the 3-axis denotes the original direction of the light beam.  
For low energy tachyons, the $T_{33}$ could be quite large.

\section {Summary}

There are several new results noted in this paper about how
faster-than-light particles (tachyons) would behave under the General 
Theory of Relativity.

First is the form of the energy-momentum tensor $T^{\mu \nu}$, where 
the spatial components would be larger than the time component. This 
is in
contrast with the situation for ordinary matter; and it raises the 
question of how this might effect theoretical models of the overall structure 
of the universe, as in the Robertson-Walker metric. 

Second is the attractive gravitational force between tachyons and the 
suggestion of a  
stable model for a closed stream of such particles. This leads to 
questions about how such structures, if they exist, might relate to  
 astrophysical observations of significance in the context of current 
cosmological models. The possibility of tachyons explaining some or 
much of what is currently called ``dark matter'' is an obvious 
thought, which I leave to better experts for consideration.

Another significant result, noted in Section 7 , is the absence of a 
long range gravitational force ($\sim 1/r^{2}$) produced by localized static 
bundles of low energy tachyons. This prediction is something that may 
be readily  subjected to observational verifiction, to prove or 
disprove the substantial existence of such tachyons.  A recent study 
\cite{KK} of tidal properties of satellite galaxies may be relevant.

In Appendix B we have noted a few additional aspects of tachyon kinematics that 
might lead to observational tests of whether or not these things 
actually exist.

This paper has been about classical particles that move faster than 
light, in the context of 
General Relativity.  Further theoretical  study is needed to look into tachyonic 
fields, how they propagate and how they may be quantized. 

\vskip 0.5cm 
\noindent{\bf Acknowledgments} 

I am grateful to K. Bardakci, S. J. Freedman, M. B. Halpern, H. C. 
Ohanian and B. Sadoulet  for some helpful conversations.

\vskip 0.5cm
\setcounter{equation}{0}
\def\theequation{A.\arabic{equation}}
\boldmath
\noindent{\bf Appendix A:  Fear of tachyons}
\unboldmath
\vskip 0.5cm

LetÕs start by reciting a version of the most common ÒparadoxÓ in considering the 
possibility of faster-than-light particles (tachyons) within the Special 
Theory of Relativity. For further reading, see Chapter 9 of
the review paper \cite{ER1} and other references therein.

A space Ship moves away from Earth (along the positive x-axis) with velocity 
$v<1$.  At time $t_{A}$, when the Ship is at coordinate $x_{0}$, a tachyon 
is sent out from Earth 
with velocity $V > 1$ in the direction of the Ship. This is event A. That 
tachyon reaches the 
Ship (this is Event B) at a later time $t_{B}$. This is the story in the Earth 
frame of reference; and it may be summarized with the following formula.

\begin{equation}
t_{B}-t_{A} = \frac{x_{0}}{V-v}.\label{A0}
\end{equation}

What is the story as seen in the space Ship frame of reference? Here we use 
the standard formulas for a Lorentz transformation and come to the 
formula
\begin{equation}
t'_{B}-t'_{A} = \gamma(1-vV)(t_{B}-t_{A}).\label{A1} 
\end{equation}
This says that, if $vV > 1$, then the tachyon arrives at the Ship before it 
was sent from Earth Ð according to observers in the Ship frame of reference.

This is the first chapter of the ÒparadoxÓ. It is no surprise to those of us 
with some sophistication with the Special Theory of Relativity. We know 
that if two events are separated by a space-like interval, then the actual 
time sequence of those two events may be different as seen in different
Lorentz frames.

The full horror of anti-causal happenings appears when one adds a 
second chapter to this story; observers in the Ship respond to 
receiving the tachyon by sending another tachyon back to Earth, even 
faster, and this is seen to arrive back on Earth \emph{before} the 
first signal was sent out. 

In order to avoid jumping to 
the conclusion that tachyons cannot exist, we should ask what lesser 
restrictions upon them might be considered. The following is a list of
assumptions  used in the story recited above.
\begin{trivlist}	
\item a) Special Relativity is correct; 
\item b) Tachyons exist; 
\item c) Tachyons can be created and sent out at will; 
\item d) Tachyons can be detected reliably.  
\end{trivlist}
It is assumptions (c) and (d) that we should address first, before we 
 reject assumption (b).

We can suggest that if tachyons exist, they should occur as 
particle-waves of the sort we describe in the quantum theory; they 
may be emitted and absorbed by interacting with ordinary matter but 
these interactions must be very weak. This could allow for a very 
uncertain experimental set-up regarding controlled emission and 
detection (absorption) of tachyons. One would want to make a 
quantitative theory and show that the probability of observing the 
sort of nasty ``paradox'' described above is swamped by unavoidable 
uncertainty or background noise.

One simple example is the following. Suppose the emission of the 
tachyon, as described above, was governed by a wave equation and 
appeared to observer A as a wave packet such as this.
\begin{equation}
\varphi(x,t) \sim  e^{i(kx-\omega 
t)}\;e^{-\Gamma(t-x/V)}\;\theta(t-x/V),\label{A2}
\end{equation}
where $V = k/\omega$ is the group velocity. This packet has a size, 
in the original reference frame, of $\Delta x = V/\Gamma$.
If we perform the Lorentz transformation on these coordinates 
to see what this wave packet 
looks like in the frame of observer B, then we find that the size of 
the wave packet is $\Delta x' = V/[\gamma \Gamma (1-vV)]$. Thus, when we 
approach that critical condition when $vV \approx 1$, the wave packet 
is so large that it encompasses both the ``emitter'' and the 
``receiver'' at the same time.  This gives us a view of how wave 
properties for tachyons can indeed erase the distinction between 
``emission'' and ``absorption'' of a particle that travels faster 
than light.  There is no paradox here.

One common idea is that 
neutrinos might be a candidate for tachyonity: they have weak 
interactions and are nevertheless plentiful throughout the universe. 
Against this idea, one might argue that neutrinos cannot be restricted by disallowing 
assumption (c) and/or (d), above, because of the possibility 
(in fact an established experiment) of using an accelerator to create 
a beam of neutrinos, which can be pulsed at the emitter and then 
detected reliably at a distant absorber. To refute this objection, 
one should notice a particular condition for the ``paradoxical'' 
situations discussed earlier. In order to get the effect of 
time-sequence reversal, between the emission and the absorption of a 
tachyon, one must have the source and the receiver moving \emph{away} from 
each other.  If they are moving toward each other, then there is no 
such effect predicted by the Lorentz transformation. (Notice what 
happens if you change the sign of $v$ in Eq. (\ref{A1}).) The arrangement 
of an accelerator sending out a large pulse of particles in the 
forward direction of the high energy beam involves just this second 
situation: source and receiver approaching one another; so there is 
no paradox to be expected here.

Of course, other arguments may be 
advanced and need to be studied.  But, at least at present, I believe this is
a constructive 
step in clarifying this controversial topic. Tachyons may exist and should not be 
dismissed from consideration in either theoretical or experimental 
studies.

\vskip 0.5cm
\setcounter{equation}{0}
\def\theequation{B.\arabic{equation}}
\boldmath
\noindent{\bf Appendix B: Kinematical musings}
\unboldmath
\vskip 0.5cm

If we calculate the number of free tachyon states that are ``on the energy 
shell'' inside a spatial volume V, we get
\begin{equation}
\sum_{states} \delta(E-E_{0}) = 
\frac{V}{2\pi^{2}}\;E_{0}\;p_{0}\label{B1}
\end{equation}
which is exactly the same formula one has for  normal 
(slower-than-light) particles.

So, if one wanted to ask whether neutrinos might actually be 
tachyons, then the search for neutrino mass, as carried out by looking 
at the end point of electron energy spectrum in beta-decay, takes on 
new possibilities.  The phase space factor (\ref{B1}) then gives us 
the spectrum for electron energies $E_{e}$:
\begin{eqnarray}
(E_{0}-E_{e})^{2} \;\;\; for\; massless\; neutrinos \\ 
(E_{0}-E_{e}) \sqrt{(E_{0}-E_{e})^{2}-m^{2}}\;\;\; for\; ordinary \;
massive\; neutrinos \\
(E_{0}-E_{e}) \sqrt{(E_{0}-E_{e})^{2}+m^{2}}\;\;\; for\; tachyon \;
 neutrinos \label{B2}
\end{eqnarray}
The second version shifts the end point of the energy spectrum by 
$m$ while the third one does not shift it. But each of the three versions 
has its own distinctive shape near the end point: The first comes in 
with zero slope; the second with infinite slope; and the third with 
finite slope. Earlier theoretical studies, as well as some experimental results,
which did not  
survive, led to the hypothesis of tachyonic neutrinos.\cite{ER4, CR}

An intriguing aspect is that the Dirac equation for tachyons gives 
special prominence to the helicity of particle states.

Also, regarding neutrinos, there are the experiments that show the 
effects of mass-mixing.  The standard view is based upon the 
expansion of the wavefunction in terms of a small mass 
\begin{equation}
e^{i(kx-\omega t)} \sim e^{i(k(x-t) - (m^{2}/2k)t)},\label{B3}
\end{equation}
and then subsequent treatment of $m^{2}$ as a matrix that causes 
mixing with other neutrino states. 

For tachyons there would be only the slightest difference:
\begin{equation}
e^{i(kx-\omega t)} \sim e^{i(\omega(x-t) + 
(m^{2}/2\omega)x)},\label{B4}
\end{equation}
and since these are high energy particles, $\omega \approx k$ and $ x 
\approx t$, it would look very much the same.

One can also ask how the simplest model for cosmological expansion, 
with uniform distribution of matter, would look in the case of a 
tachyon-dominated universe.  The basic equations are
\begin{eqnarray}
\frac{d}{da}(\rho a^{3}) = -3 P a^{2}\label{B5} \\ 
\rho = const. \times \int E\;d^{3}p/(e^{E/kT} \pm 1)\label{B6} ,
\end{eqnarray}
where $a$ is the scale factor and $T$ is the temperature, with $\rho, 
P$ being the energy density and pressure as previously defined.  In 
the case where $kT >> mc^{2}$, where $m$ is the tachyonic mass, then 
we have the familiar results for any relativistic particle: $\rho = 
3P  
\sim T^{4} \sim a^{-4}$. Now we want to see what this looks like in 
the other limit, when $kT << mc^{2}$. Then we find that $ \rho \sim 
T^{3}$ and, if we assume Fermi statistics, then the pressure $P$ goes 
as $T$. This leads to a relation between $T$ and $a$ as follows:
\begin{equation}
\left(\frac{T}{T_{1}}\right)^{2} = 2 \left(\frac{a_{1}}{a}\right)^{2} - 1.\label{B7}
\end{equation}
This is a rather unusual result, predicting that the temperature goes 
to zero at some finite expansion. Of course, this very simple-minded 
calculation needs to be studied further; and it might lead to some 
observational comparisons that help us decide whether such tachyons 
exist. For some related investigations, see \cite{KGKGP}.

\vskip 0.5cm
\setcounter{equation}{0}
\def\theequation{C.\arabic{equation}}
\boldmath
\noindent{\bf Appendix C:  Smoothed cylindrical problem}
\unboldmath
\vskip 0.5cm

In order to arrive at a clear solution of the field equations for 
the static-cylindrical problem studied in Section 5, we need to specify some structure 
for the source.  We shall do this in a backwards manner: make up a 
solution for the metric and see what source it produces from the 
Einstein field equations.  We start with a modified version of the 
solutions given in the earlier Section.
\begin{eqnarray}
A = a s^{\alpha}, \;\;\; B = b s^{\beta}, \;\;\; C = c 
s^{\gamma}, \;\;\; s = \sqrt{r^{2} + \epsilon^{2}}\\ 
\beta = \alpha^{2}/(\alpha+2),\;\;\;\gamma = 
-2\alpha/(\alpha+2).\label{C1}
\end{eqnarray}

We then calculate,
\begin{eqnarray}
G_{00} = - \frac{A}{2B} 
\;\frac{\alpha(\alpha+4)}{\alpha+2}\;\frac{\epsilon^{2}}{s^{4}},
\;\;\;\;\; G_{33} = \frac{C}{A}\;G_{00} \\
G_{11} =- \frac{1}{2}\;\frac{\alpha^{2}}{\alpha+2}\;
\frac{\epsilon^{2}}{s^{4}}, \;\;\;\;\; 
G_{22} = \frac{r^{2}}{B}\;G_{11} .\label{C2}
\end {eqnarray}
Next, we integrate these densities $\int_{0}^{\infty}r\;dr$ and find, 
in the limit of small $\alpha$ and small $\epsilon$:
\begin{eqnarray}
\int G_{00} \longrightarrow -\;\frac{a}{2b}\;\alpha = -4G \rho \gamma\\ 
\int G_{33} \longrightarrow -\;\frac{c}{2b}\;\alpha = -4G\rho  \gamma 
v^{2}.\label{C3}
\end{eqnarray}

This verifies the results given earlier.

\end{document}